%% file: 26ISCAS_EZ130.tex
\documentclass[conference]{IEEEtran}

\IEEEoverridecommandlockouts

\usepackage{amsmath,amsfonts}
\usepackage{amssymb, amsthm}
\usepackage{graphicx}
\usepackage{subfigure}
\usepackage{booktabs}
\usepackage{cite}
\usepackage{xcolor}
\usepackage{microtype}
\usepackage{hyperref}
\usepackage{balance}

\allowdisplaybreaks

\input{macros/vmr-symbols-vecbold}
\input{macros/standard-macros}

\input{macros/defs}

\renewcommand{\bml}{\ensuremath{\boldsymbol \ell}}

\begin{document}
\title{An Open-Source Standard-Cell Library \\ for IHP 130\,nm Developed by Students}

\author{
\IEEEauthorblockN{Oscar Casta\~neda$^1$, Lavinia Recchioni$^2$, Tobias Senti$^3$, Flurin Cahenzli$^3$, Marco Ferroni$^3$, Thorben Heekenjann$^3$, \\ Robert Kenter$^3$, Stefan Odermatt$^3$, Daniel Richner$^3$, Gabriel Altendorfer$^3$, Davide Cannone$^3$, Miguel Correa$^3$, \\ Ivan Herger$^3$, Lars Kr\"oger$^3$, Nicolas Nanzer$^3$, Darja Nonaca$^3$, Christopher Reinwardt$^3$, Lukas Winklhofer$^3$, \\ Enrico Zelioli$^3$, Yiheng Zhang$^3$, Yingxue Zhang$^3$, Domenic Keller$^4$, Seyed Hadi Mirfarshbafan$^4$, Zerun Jiang$^4$,\\ Beat Muheim$^4$, Arianna Rubino$^4$, J\'er\'emy Guichemerre$^4$, Frank K. G\"urkaynak$^1$, and Christoph Studer$^1$} \\[-0.3cm]
\IEEEauthorblockA{\emph{Department of Information Technology and Electrical Engineering, ETH Zurich, Switzerland}}
\thanks{The authors' superscripts indicate their roles in the VLSI~5 course: $^1$lecturer, $^2$teaching assistant, $^3$student (ordered by number of contributed standard cells and alphabetically in case of a tie), and $^4$course-setup contributor.}\thanks{This work has received funding from the Swiss State Secretariat for Education, Research, and Innovation (SERI) under the SwissChips initiative.}\thanks{The authors thank Cadence for their support by providing licenses to design and characterize the EZ library. The authors also thank H. Pretl, T. Benz, and P. Sauter for their feedback on preliminary versions of the EZ library.}\thanks{Contact author: O. Casta\~neda (e-mail: caoscar@ethz.ch)}
}
\maketitle

\begin{abstract} 
Standard-cell libraries form the critical interface between transistor-level circuit design and automated digital design flows, yet their design and characterization are rarely covered in depth in digital VLSI courses. This paper presents VLSI 5, a graduate-level course at ETH Zurich in which students design, lay out, characterize, and integrate their own standard cells using IHP's open-source 130\,nm SG13G2 process design kit. In the first course offering, 19 students developed EZ130 8T, an open-source eight-track standard-cell library comprising 220 cells. We evaluate successive library versions using eight synthesis benchmarks, across which EZ130 8T achieves significant area reduction compared to the SG13G2 library at competitive timing. For a postlayout 8$\times$8 matrix-vector multiplier, EZ130 8T reduces cell area by 35\% and energy by 39\% while maintaining virtually the same clock period. The library has already been adopted in other ETH Zurich VLSI courses, enabling the creation of chips entirely designed by students down to the transistor level.
\end{abstract}

\section{Introduction}

Modern digital integrated-circuit (IC) design relies on abstraction layers that separate transistor-level circuits from high-level hardware descriptions. Standard cells form a critical interface between these levels: digital designers typically treat them as pre-designed components, while decisions such as transistor sizing, cell height, and the tradeoffs among area, delay, and energy are left to specialized cell designers.

While digital VLSI courses commonly introduce transistor-level design and layout of basic standard cells, few give students the opportunity to cross this abstraction boundary by designing, characterizing, and integrating their own cells into a digital IC design flow. One notable exception is the course described in~\cite{stine03}, in which students designed and characterized a standard-cell library using a proprietary process design kit (PDK). Such proprietary PDKs, however, limit the sharing and reproduction of course infrastructure and the student-created library. 

Open PDKs provide an opportunity to overcome these limitations by making the resulting design artifacts accessible and reproducible. While widely used open PDKs such as FreePDK~\cite{freepdk07} and ASAP7~\cite{asap7} describe virtual technologies that cannot be fabricated, more recent PDKs target physical foundry processes, including SkyWater's SKY130~\cite{sky130}, GlobalFoundries' GF180MCU~\cite{gf180mcu}, and IHP's SG13G2~\cite{ihp13}. These PDKs enable educational projects whose results can not only be shared and reproduced, but also fabricated and measured.

\subsubsection*{Contributions} 
We present VLSI 5, a graduate-level course at ETH Zurich in which students design, lay out, and characterize their own standard cells and integrate them into a digital IC design flow. In its first offering, 19 students developed EZ130 8T, an eight-track open-source standard-cell library with 220 cells for IHP's SG13G2~\cite{ihp13}, an open-source 130\,nm technology.
We describe the course methodology and evaluate the resulting library through synthesis and place-and-route benchmarks.
\setlength{\fboxsep}{6pt}   
\setlength{\fboxrule}{0.5pt} 
\begin{center}
\fbox{%
  \begin{minipage}{0.8\columnwidth}
    \centering
 The open-source EZ130 8T standard-cell \\ library is available at \url{https://ez.ethz.ch} 
  \end{minipage}%
}
\end{center}

\section{VLSI~5: A Course on Standard-Cell Design}

\subsection{Course Context}

\begin{figure}[t]
\centering
\includegraphics[width=0.99\linewidth]{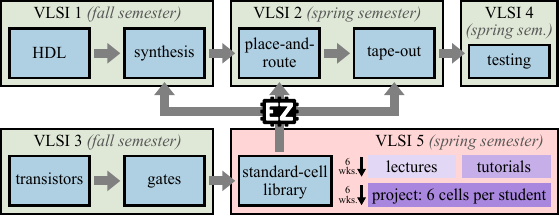}
  \caption{ETH Zurich course series on digital VLSI design. By teaching students to create a standard-cell library, the new VLSI 5 course closes the gap between transistor-level and HDL-level IC design.}
  \label{fig:courses}
\end{figure}

\begin{figure*}[t]
  \begin{minipage}{0.6\linewidth}
    \centering
    \refstepcounter{table}
    \label{tab:topics}
    \footnotesize
    \text{TABLE~\thetable}\\
    \textsc{Topics Covered by the ``VLSI 5: Design of a Standard-Cell Library'' Course}\\
    ~\\[-0.1cm]
    \scalebox{0.99}{
    \begin{tabular}{@{}cll@{}}
      \toprule
      Week & Theoretical lecture & Practical tutorial \\
      \midrule
      1 & Principles of standard-cell design & Exploring the IHP 130\,nm technology \\
      2 & Modeling of CMOS gates & Manual characterization of gates and SPICE \\
      \multirow[t]{2}{*}{3} & Functional characterization of & Cadence Liberate, LIB and DB files,\\
         & gates and synthesis & and Synopsys DC\\
      \multirow[t]{2}{*}{4} & Physical characterization of & \multirow[t]{2}{*}{LEF files and Cadence Innovus}\\
         & gates and back-end design & \\
      5 & Design verification & CDL files, Siemens Calibre and QuestaSim \\
      6 & Tips and tricks for layout & Project organization \\
      \bottomrule
    \end{tabular}
    }  
  \end{minipage}
  \hfill
  \begin{minipage}{0.38\linewidth}    
    \centering
    \includegraphics[width=0.8\linewidth]{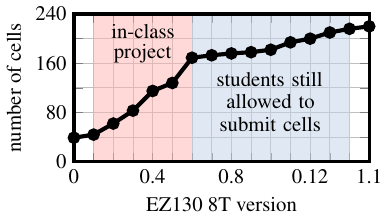}
    \caption{Number of cells per EZ130 8T library version.}
    \label{fig:cellprogress}
  \end{minipage}
\end{figure*}

ETH Zurich has traditionally followed a top-down approach for teaching digital VLSI design. 
Until 2020, three digital VLSI courses were offered: VLSI~1 teaches front-end design using hardware description languages (HDLs) and synthesis; VLSI~2 teaches back-end design, from floorplanning through placement-and-routing to layout verification; and VLSI~4\footnote{The original name of VLSI~4 was VLSI~III, but it was renamed in 2021.} teaches the practical aspects of IC testing.
Together, these courses prepare students to design their own ICs, send them to fabrication, and even test them after fabrication.

This top-down approach, however, was not able to fully delve into the working principles of MOSFETs and how they are used to build logic gates, which would occasionally lead to misunderstandings when designing or testing ICs.
To fill this gap, a fourth course, VLSI~3, was introduced in 2021. 
VLSI~3 was well received by the students and has experienced a 150\% increase in enrollment from 2021 to 2025. 
While VLSI~3 covered the foundations of MOSFETs, CMOS, standard-cell design, and digital layout drawing, there was still no course in which the students would learn to interface their low-level transistor-level designs with their high-level HDL designs, as illustrated by \fref{fig:courses}.
Therefore, in 2025, we designed a new course, ``VLSI~5: Design of a Standard-Cell Library,'' completing a series of courses that provide a holistic perspective on digital IC design---all the way from transistor to chip.

\subsection{Course Organization}

In VLSI~5, students at the Master's level learn to design and characterize their own custom standard cells, and to generate all associated interface files (LIB, LEF, Verilog, SPICE/CDL, and GDSII), so that their cells can be integrated into a standard digital IC design flow. 
The course consists of a weekly 3-hour session over 12 weeks, with the students also working outside class hours for a total of 4 ECTS (European Credit Transfer and Accumulation System) credits. 
VLSI 5 is an ungraded course, which means that the students can either pass or fail.

As shown in \fref{fig:courses}, VLSI 5 consists of three components: theoretical lectures, hands-on tutorials, and an in-semester project, in which the students solidify their learning by doing.
The lectures and tutorials take place during the first six weeks of the semester and go hand in hand: between one and two hours of each weekly session are used to cover a topic theoretically, and the remaining time is used to follow a practical tutorial on the same topic. 
Table~\ref{tab:topics} lists the topics covered during this phase of the course. 
The tutorials teach the students how to generate all interface files from the schematic and layout of a standard cell, and how to integrate such a cell into a digital IC design flow. 
The weekly sessions of the last six weeks are dedicated to the course's project, which is described next.

\subsection{Course Project}

To pass the course, each student had to create six new standard cells.
Creating a new standard cell entails (i) drawing the schematic and layout with Cadence Virtuoso, (ii) performing functional characterization (LIB and Verilog files) with Cadence Liberate and physical characterization (LEF file) with Cadence Abstract Generator, as well as generating the CDL and GDSII files, and (iii) implementing a digital design that instantiates the new standard cell to verify that the cell integrates correctly within a digital IC design flow. 
For the digital IC design flow, we used Synopsys DC for synthesis, Cadence Innovus for back-end design, Siemens Calibre for physical verification, and Siemens QuestaSim for functional verification.
We used commercial CAD tools to broaden the students' toolkit, as they already use open-source CAD tools in the VLSI~2 course.

To provide the students with a starting point, we created a seed standard-cell library with 39 cells as well as layout templates. 
We also provided the students with reference HDL designs and with scripts that largely automate the digital IC design flow, so that they could focus on the cells' design process and not be hindered by tool complexity. 
Since each student had to implement six cells distinct from those assigned to their peers, personalized feedback was needed. 
Thus, we opted for an in-class project so that the students could receive prompt in-person assistance. 
The students could submit their completed cells every week, which were then evaluated by the course's instructors and either accepted as final or sent back for corrections.
\fref{fig:cellprogress} shows the progress of accepted cells throughout the semester: the first six weekly submissions occurred during the lecture period of the semester; students could still submit cells afterwards.

At the beginning of the course, most of the cell evaluation was performed manually by the instructors.
This quickly proved to be time consuming, so, throughout the semester, scripts and custom design-rule-check (DRC) decks were created to automate most of these checks.
Once a student got all of their six cells satisfying all checks, they would also create a datasheet for their cells.
Combining all of these datasheets resulted in the datasheet for the complete standard-cell library.

\subsection{Course Results}
The first offering of the VLSI~5 course had 19 students; the number of students was limited given that it was the first offering of the course and the highly personalized feedback it required.
Together, the students created 164 cells, which is more than the mandatory 114 cells, as several students were highly motivated and created additional cells (a single student even created 40 additional cells). 
All of these cells have been released---with the students' consent---as an open standard-cell library called EZ (short for ETH Zurich; see \fref{sec:ez}).
The course received positive evaluations, with a general satisfaction score of 4.7 out of 5.0, far above the department average.


\section{The EZ130 8T Open Standard-Cell Library}
\label{sec:ez}

\subsection{Technical Specifications}

\begin{figure}[t]
\centering
\subfigure[AND3X2]{\includegraphics[width=.21\columnwidth]{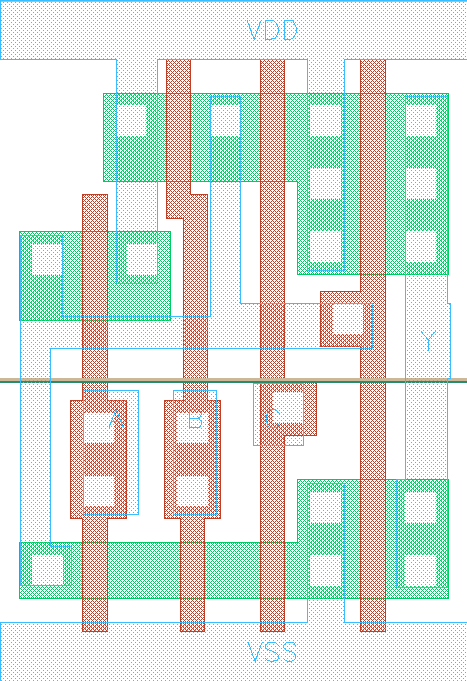}\label{fig:cell_and3}}
\hfill
\subfigure[D-type flip-flop DFFRQX2]{\includegraphics[width=.70\columnwidth]{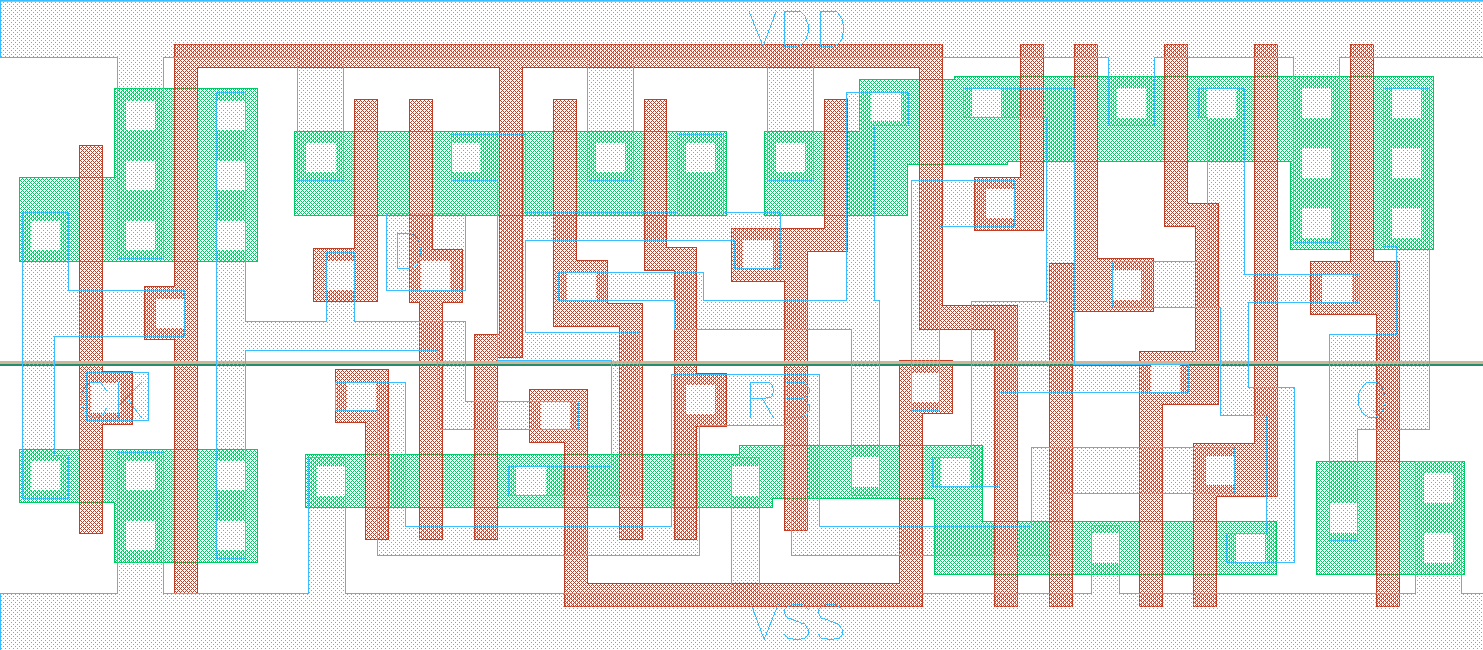}\label{fig:cell_dff}}
\caption{Layout of two example standard cells from the EZ130 8T library.}\label{fig:cells}
\end{figure}

\begin{figure*}[tp]
\centering
\subfigure[32-bit adder]{\includegraphics[width=.48\columnwidth]{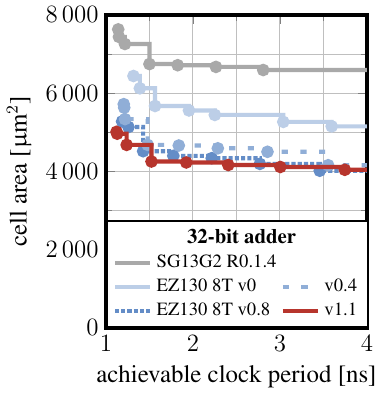}\label{fig:synth_add}}
\hfill
\subfigure[32-bit multiplier]{\includegraphics[width=.48\columnwidth]{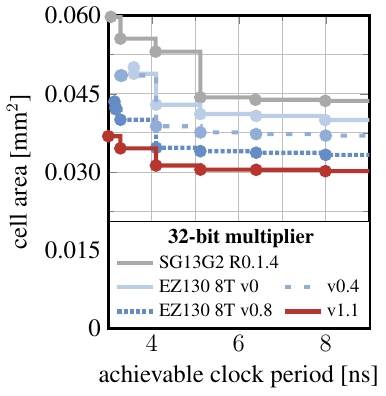}\label{fig:synth_mult}}
\hfill
\subfigure[24-bit counter]{\includegraphics[width=.48\columnwidth]{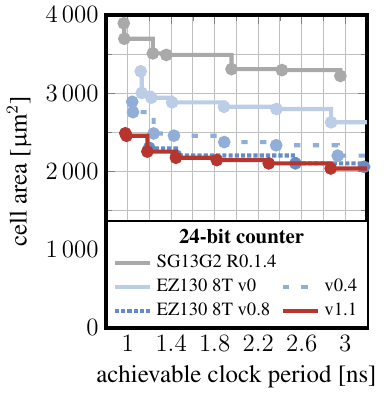}\label{fig:synth_counter}}
\hfill
\subfigure[8b-input 8b-output LUT]{\includegraphics[width=.48\columnwidth]{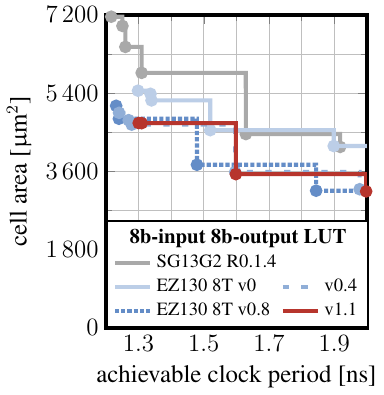}\label{fig:synth_lut}}\\
\subfigure[10-stage 24-bit shift register]{\includegraphics[width=.48\columnwidth]{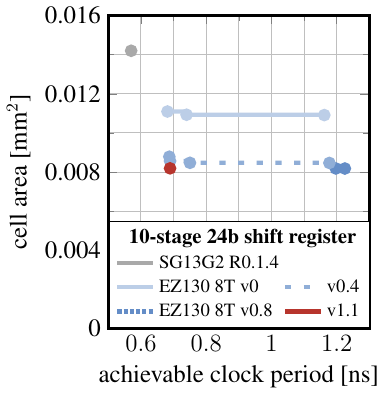}\label{fig:synth_shift}}
\hfill
\subfigure[32$\times$32-bit latch array]{\includegraphics[width=.48\columnwidth]{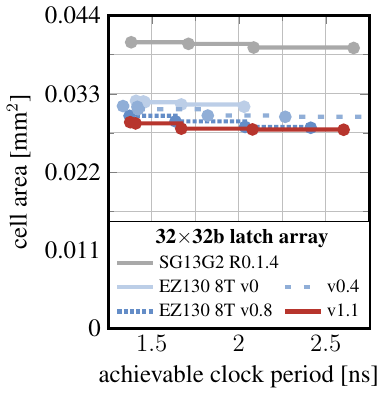}\label{fig:synth_latch}}
\hfill
\subfigure[64-point FFT]{\includegraphics[width=.465\columnwidth]{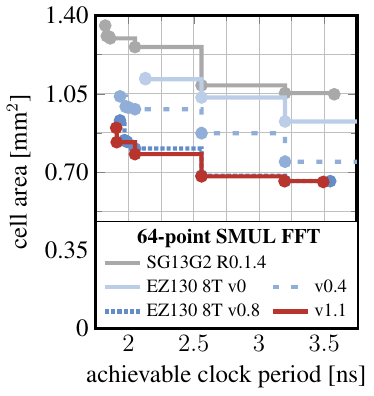}\label{fig:synth_fft}}
\hfill
\subfigure[8$\times$8 matrix-vector multiplier]{\includegraphics[width=.465\columnwidth]{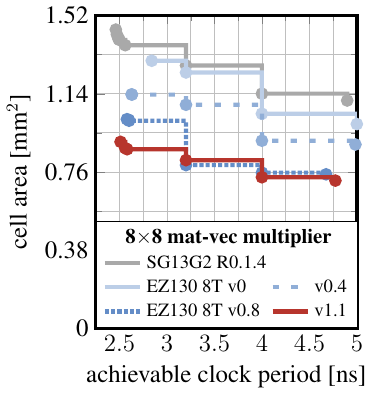}\label{fig:synth_lut}}
\caption{Synthesis results for the EZ130 8T and SG13G2 libraries obtained for eight different benchmark designs while sweeping the clock period constraint.}\label{fig:synth_all}
\end{figure*}

The EZ130 8T library combines the 39 cells from the VLSI~5 seed library with the 164 cells created by the students and additional cells created by the instructors, resulting in a total of 220 cells.
All these cells were implemented with SG13G2, the 130\,nm BiCMOS open-source PDK from IHP~\cite{ihp13}.
The cells were designed for a routing grid of 0.42\,$\upmu$m x 0.42\,$\upmu$m.
To facilitate automated routing, the routing grid for Metal~1 and Metal~3 are aligned and all cell pins are placed on grid points.
As its name indicates, the cells in EZ130 8T were designed for a height of eight tracks (3.36\,$\upmu$m).
Only Metal~1 is used for routing within the cells, and is typically routed at its minimum width of 0.16\,$\upmu$m; power rails are routed with a width of 0.32\,$\upmu$m.
Well taps are not integrated within each cell beneath its power rails but are implemented as dedicated cells.

The cells are available in different driving strengths, mostly X1 and X2.
For X2 cells, the NMOS and PMOS widths are 0.64 and 0.98\,$\upmu$m, respectively, corresponding to the minimum widths fitting two and three diffusion contacts.
Buffers and inverters are available with a driving strength of up to X64.
The EZ130 8T library contains combinational and sequential logic cells, as well as physical cells (e.g., fillers with decoupling, antenna diode, and tie cells). 
The combinational cells cover all possible transistor topologies for monolithic CMOS gates with up to four inputs.
The library also includes non-monolithic gates (e.g., AND3 and OR2B), tristate cells, flip-flops, latches, and clock gates.
\fref{fig:cells} shows two example EZ130 8T cells.

\subsection{Benchmarks}
\label{sec:benchmarks}

To evaluate the progress of the EZ library as more cells were added throughout the semester and to assess the library's overall performance, we used several benchmark HDL designs to synthesize and place-and-route with our cells.
These benchmark designs had to be simple enough to allow quick evaluation, while also covering a meaningful and broad range of use cases.
The benchmarks we used are as follows:
\begin{itemize}
\item \emph{Adder:} An adder with two 32-bit inputs and one 32-bit output (i.e., the carry-out bit is discarded). Both inputs and the output are registered with flip-flops.
\item \emph{Multiplier:} A signed multiplier with two 32-bit inputs and one 64-bit output, which are all registered with flip-flops.
\item \emph{Counter:} An adder increments a 24-bit register by one on every clock cycle. The count can be disabled or reset.
\item \emph{Look-up table (LUT):} A randomly generated LUT with 256 8-bit entries, synthesized as random combinational logic. The inputs and outputs of this LUT are registered.
\item \emph{Shift register:} Ten 24-bit registers connected in series.
\item \emph{Latch array:} A random-access memory storing 32 words of 32 bits each using latches, following~\cite{meinerzhagen2010}. The inputs and outputs of this memory are registered with flip-flops.
\end{itemize}

In addition, we used two more complex benchmark designs, which, although slower to implement, allowed us to evaluate the library in more comprehensive applications:
\begin{itemize}
\item \emph{64-point fast Fourier transform (FFT):} A 64-point \mbox{radix-4} FFT with complex-valued 7-bit inputs, implemented using the fully unrolled, multiplierless architecture from \cite{mirfarshbafan21}.
\item \emph{8$\times$8 matrix-vector multiplier:} Eight instances of a complex-valued inner-product engine~\cite{mirfarshbafan26}, that pointwise multiplies two complex-valued vectors (each with eight 8-bit entries) and then sums up the partial products. The design is fully unrolled, effectively computing one 8$\times$8 matrix-vector product (MVP) per clock cycle.
\end{itemize}

\subsection{Synthesis Results}

\fref{fig:synth_all} shows area vs. achievable clock period plots for the different benchmarks from \fref{sec:benchmarks}.
To generate these plots, we performed a first synthesis with Synopsys DC using the nominal TT, 1.20\,V, 25$^\circ$C corner and a relaxed clock constraint.
The clock constraint of the next synthesis was then adjusted based on the slack of the previous synthesis.
This process was repeated until we observed no significant improvement in the achievable clock period (i.e., clock constraint minus slack).
We then extracted the Pareto-optimal envelope of the synthesized designs, i.e., those designs that achieve the lowest area at a given clock period or the lowest clock period at a given area.

\fref{fig:synth_all} compares several versions of the EZ130 8T library.
Version 0 (v0) is the 39-cell seed library from VLSI~5, while subsequent versions accumulate the cells accepted each week of the course's project; for example, v0.4 contains all cells accepted through the fourth project week.
Version 1.0 marks the end of the semester, and subsequent lecturer optimizations resulted in the current EZ130 8T v1.1.
As a baseline, we include synthesis results for revision 0.1.4 of the standard-cell library distributed with the SG13G2 PDK~\cite{ihp13}.
This SG13G2 library contains 84 cells with a 3.78\,$\upmu$m height and integrated well taps. The SG13G2 library has mostly X1 and X2 driving strengths, with X2 cells using NMOS and PMOS widths of 1.48 and 2.24\,$\upmu$m, respectively, 2.3$\times$ those of EZ130 8T.
For a fair comparison, we re-characterized the SG13G2 library with our Cadence Liberate flow, while preserving the indices of its nonlinear model's LUTs.

\fref{fig:synth_all} shows that the initial EZ130 8T v0 library is already smaller than SG13G2 after synthesis, albeit slower.
Each version of the EZ130 8T library becomes smaller and overall faster, to the point that EZ130 8T v1.1 achieves a clock period within 120\,ps of that of SG13G2 for all benchmarks.

\subsection{Place-and-Route Results}

\begin{table}[t]
\setlength{\tabcolsep}{4pt} 
\caption{Place-and-Route Results for 8$\times$8 matrix-vector multiplier}
\label{tab:pnr_results}
\vspace{-0.2cm}
\centering
\scalebox{0.99}{
\begin{tabular}{@{}lccc@{}}
  \toprule
  \multirow{3}{*}{Library} & SG13G2 & \multicolumn{2}{c}{~~EZ130 8T} \\
  \cmidrule(){2-2}\cmidrule(l){3-4}
   & R0.1.4 & v0 & v1.1 \\
  \midrule
  Cell area $A$ [mm$^2$] & 1.42 & 1.36 & \textbf{0.92} \\
  Min. clock period $T$ [ns] & \textbf{4.89} & 6.05 & 5.05 \\
  Energy $E$ [nJ/MVP] & 1.27 & 0.97 & \textbf{0.77} \\
  $A\times T\times E$ [normalized] & 1.00 & 0.90 & \textbf{0.41} \\ 
  \bottomrule
\end{tabular}
}
\end{table}

For our most complex benchmark, the 8$\times$8 complex-valued matrix-vector multiplier, we performed back-end design of the synthesized netlist with the lowest area-delay product using Cadence Innovus. We swept the clock constraint in 100\,ps steps, while targeting 50\% density and analyzing three corners: (i)~TT, 1.20\,V, 25\,$^\circ$C; (ii) FF, 1.32\,V, -40\,$^\circ$C; and (iii) SS, 1.08\,V, 125\,$^\circ$C.
Table~\ref{tab:pnr_results} shows the post-layout results with the best area-delay product per library. The reported clock period is met across all corners. Energy is evaluated at the TT corner and the reported clock period using stimuli from MVPs between a fixed randomly generated matrix and 100 random input vectors.

Similar to the synthesis results, EZ130 8T v0 is smaller than SG13G2 but significantly slower, while EZ130 8T v1.1 reduces area by 35\% and energy consumption by 39\% at a 3\% larger clock period---we attribute these savings to the use of smaller transistors. 
Altogether, the EZ130 8T library offers a 59\% smaller area-delay-energy product than SG13G2.

\section{Conclusions}
We have created a new course at ETH Zurich: ``VLSI~5: Design of a Standard-Cell Library'' that teaches students to design their own custom cells and to integrate them into a digital IC design flow.
VLSI~5 completes a series of digital VLSI courses through which students gain a holistic perspective on digital IC design, from understanding the physical devices that constitute ICs to designing, implementing, and testing their own IC and the standard cells that form it.
As part of the VLSI~5 project, the students developed an open-source standard-cell library, EZ130 8T, for the open-source 130\,nm PDK from IHP, SG13G2. 
Post-layout results for a matrix-vector multiplier show that EZ130 achieves 35\% smaller area and 39\% lower energy than the SG13G2 library, at the same speed.

The VLSI~5 course was well received by the students and was offered again in spring 2026. The EZ130~8T v1.0 library was adopted in the 2026 offering of VLSI~2, where students implemented and taped out ICs using cells created by their peers. Thus, every component in these ICs was designed by students.
Two chips with EZ130 8T cells, one with v0 and one with v0.12, have come back from fabrication and have passed initial tests; their complete characterization is ongoing work.

\end{document}

%% file: macros/vmr-symbols-vecbold.tex
\usepackage{amssymb}
\usepackage{amsfonts}
\usepackage{mathrsfs}
\usepackage{xspace}
\usepackage{bm}
\usepackage{upgreek}

\newcommand{\safemath}[2]{\newcommand{#1}{\ensuremath{#2}\xspace}}

\safemath{\bma}{\mathbf{a}}
\safemath{\bmb}{\mathbf{b}}
\safemath{\bmc}{\mathbf{c}}
\safemath{\bmd}{\mathbf{d}}
\safemath{\bme}{\mathbf{e}}
\safemath{\bmf}{\mathbf{f}}
\safemath{\bmg}{\mathbf{g}}
\safemath{\bmh}{\mathbf{h}}
\safemath{\bmi}{\mathbf{i}}
\safemath{\bmj}{\mathbf{j}}
\safemath{\bmk}{\mathbf{k}}
\safemath{\bml}{\mathbf{l}}
\safemath{\bmm}{\mathbf{m}}
\safemath{\bmn}{\mathbf{n}}
\safemath{\bmo}{\mathbf{o}}
\safemath{\bmp}{\mathbf{p}}
\safemath{\bmq}{\mathbf{q}}
\safemath{\bmr}{\mathbf{r}}
\safemath{\bms}{\mathbf{s}}
\safemath{\bmt}{\mathbf{t}}
\safemath{\bmu}{\mathbf{u}}
\safemath{\bmv}{\mathbf{v}}
\safemath{\bmw}{\mathbf{w}}
\safemath{\bmx}{\mathbf{x}}
\safemath{\bmy}{\mathbf{y}}
\safemath{\bmz}{\mathbf{z}}
\safemath{\bmzero}{\mathbf{0}}
\safemath{\bmone}{\mathbf{1}}

\bmdefine{\biad}{a}
\bmdefine{\bibd}{b}
\bmdefine{\bicd}{c}
\bmdefine{\bidd}{d}
\bmdefine{\bied}{e}
\bmdefine{\bifd}{f}
\bmdefine{\bigd}{g}
\bmdefine{\bihd}{h}
\bmdefine{\biid}{i}
\bmdefine{\bijd}{j}
\bmdefine{\bikd}{k}
\bmdefine{\bild}{l}
\bmdefine{\bimd}{m}
\bmdefine{\bind}{n}
\bmdefine{\biod}{o}
\bmdefine{\bipd}{p}
\bmdefine{\biqd}{q}
\bmdefine{\bird}{r}
\bmdefine{\bisd}{s}
\bmdefine{\bitd}{t}
\bmdefine{\biud}{u}
\bmdefine{\bivd}{v}
\bmdefine{\biwd}{w}
\bmdefine{\bixd}{x}
\bmdefine{\biyd}{y}
\bmdefine{\bizd}{z}

\bmdefine{\bixid}{\xi}
\bmdefine{\bilambdad}{\lambda}
\bmdefine{\bimud}{\mu}
\bmdefine{\bithetad}{\theta}
\bmdefine{\biphid}{\phi}
\bmdefine{\bideltad}{\delta}

\safemath{\bmia}{\biad}
\safemath{\bmib}{\bibd}
\safemath{\bmic}{\bicd}
\safemath{\bmid}{\bidd}
\safemath{\bmie}{\bied}
\safemath{\bmif}{\bifd}
\safemath{\bmig}{\bigd}
\safemath{\bmih}{\bihd}
\safemath{\bmii}{\biid}
\safemath{\bmij}{\bijd}
\safemath{\bmik}{\bikd}
\safemath{\bmil}{\bild}
\safemath{\bmim}{\bimd}
\safemath{\bmin}{\bind}
\safemath{\bmio}{\biod}
\safemath{\bmip}{\bipd}
\safemath{\bmiq}{\biqd}
\safemath{\bmir}{\bird}
\safemath{\bmis}{\bisd}
\safemath{\bmit}{\bitd}
\safemath{\bmiu}{\biud}
\safemath{\bmiv}{\bivd}
\safemath{\bmiw}{\biwd}
\safemath{\bmix}{\bixd}
\safemath{\bmiy}{\biyd}
\safemath{\bmiz}{\bizd}

\safemath{\bmxi}{\bixid}
\safemath{\bmlambda}{\bilambdad}
\safemath{\bmmu}{\bimud}
\safemath{\bmtheta}{\bithetad}
\safemath{\bmphi}{\biphid}
\safemath{\bmdelta}{\bideltad}

\safemath{\bA}{\mathbf{A}}
\safemath{\bB}{\mathbf{B}}
\safemath{\bC}{\mathbf{C}}
\safemath{\bD}{\mathbf{D}}
\safemath{\bE}{\mathbf{E}}
\safemath{\bF}{\mathbf{F}}
\safemath{\bG}{\mathbf{G}}
\safemath{\bH}{\mathbf{H}}
\safemath{\bI}{\mathbf{I}}
\safemath{\bJ}{\mathbf{J}}
\safemath{\bK}{\mathbf{K}}
\safemath{\bL}{\mathbf{L}}
\safemath{\bM}{\mathbf{M}}
\safemath{\bN}{\mathbf{N}}
\safemath{\bO}{\mathbf{O}}
\safemath{\bP}{\mathbf{P}}
\safemath{\bQ}{\mathbf{Q}}
\safemath{\bR}{\mathbf{R}}
\safemath{\bS}{\mathbf{S}}
\safemath{\bT}{\mathbf{T}}
\safemath{\bU}{\mathbf{U}}
\safemath{\bV}{\mathbf{V}}
\safemath{\bW}{\mathbf{W}}
\safemath{\bX}{\mathbf{X}}
\safemath{\bY}{\mathbf{Y}}
\safemath{\bZ}{\mathbf{Z}}

\safemath{\bZero}{\mathbf{0}}
\safemath{\bOne}{\mathbf{1}}
\safemath{\bDelta}{\mathbf{\Delta}}
\safemath{\bLambda}{\mathbf{\UpLambda}}
\safemath{\bPhi}{\mathbf{\Upphi}}
\safemath{\bSigma}{\mathbf{\Upsigma}}
\safemath{\bOmega}{\mathbf{\Upomega}}
\safemath{\bTheta}{\mathbf{\Uptheta}}

\bmdefine{\biAd}{A}
\bmdefine{\biBd}{B}
\bmdefine{\biCd}{C}
\bmdefine{\biDd}{D}
\bmdefine{\biEd}{E}
\bmdefine{\biFd}{F}
\bmdefine{\biGd}{G}
\bmdefine{\biHd}{H}
\bmdefine{\biId}{I}
\bmdefine{\biJd}{J}
\bmdefine{\biKd}{K}
\bmdefine{\biLd}{L}
\bmdefine{\biMd}{M}
\bmdefine{\biOd}{N}
\bmdefine{\biPd}{O}
\bmdefine{\biQd}{P}
\bmdefine{\biRd}{R}
\bmdefine{\biSd}{S}
\bmdefine{\biTd}{T}
\bmdefine{\biUd}{U}
\bmdefine{\biVd}{V}
\bmdefine{\biWd}{W}
\bmdefine{\biXd}{X}
\bmdefine{\biYd}{Y}
\bmdefine{\biZd}{Z}

\bmdefine{\biDelta}{\Delta}
\bmdefine{\biLambda}{\Lambda}
\bmdefine{\biPhi}{\Phi}
\bmdefine{\biSigma}{\Sigma}
\bmdefine{\biOmega}{\Omega}
\bmdefine{\biTheta}{\Theta}

\safemath{\bimA}{\biAd}
\safemath{\bimB}{\biBd}
\safemath{\bimC}{\biCd}
\safemath{\bimD}{\biDd}
\safemath{\bimE}{\biEd}
\safemath{\bimF}{\biFd}
\safemath{\bimG}{\biGd}
\safemath{\bimH}{\biHd}
\safemath{\bimI}{\biId}
\safemath{\bimJ}{\biJd}
\safemath{\bimK}{\biKd}
\safemath{\bimL}{\biLd}
\safemath{\bimM}{\biMd}
\safemath{\bimN}{\biNd}
\safemath{\bimO}{\biOd}
\safemath{\bimP}{\biPd}
\safemath{\bimQ}{\biQd}
\safemath{\bimR}{\biRd}
\safemath{\bimS}{\biSd}
\safemath{\bimT}{\biTd}
\safemath{\bimU}{\biUd}
\safemath{\bimV}{\biVd}
\safemath{\bimW}{\biWd}
\safemath{\bimX}{\biXd}
\safemath{\bimY}{\biYd}
\safemath{\bimZ}{\biZd}

\safemath{\bimDelta}{\biDelta}
\safemath{\bimLambda}{\biLambda}
\safemath{\bimPhi}{\biPhi}
\safemath{\bimSigma}{\biSigma}
\safemath{\bimOmega}{\biOmega}
\safemath{\bimTheta}{\biTheta}

\safemath{\setA}{\mathcal{A}}
\safemath{\setB}{\mathcal{B}}
\safemath{\setC}{\mathcal{C}}
\safemath{\setD}{\mathcal{D}}
\safemath{\setE}{\mathcal{E}}
\safemath{\setF}{\mathcal{F}}
\safemath{\setG}{\mathcal{G}}
\safemath{\setH}{\mathcal{H}}
\safemath{\setI}{\mathcal{I}}
\safemath{\setJ}{\mathcal{J}}
\safemath{\setK}{\mathcal{K}}
\safemath{\setL}{\mathcal{L}}
\safemath{\setM}{\mathcal{M}}
\safemath{\setN}{\mathcal{N}}
\safemath{\setO}{\mathcal{O}}
\safemath{\setP}{\mathcal{P}}
\safemath{\setQ}{\mathcal{Q}}
\safemath{\setR}{\mathcal{R}}
\safemath{\setS}{\mathcal{S}}
\safemath{\setT}{\mathcal{T}}
\safemath{\setU}{\mathcal{U}}
\safemath{\setV}{\mathcal{V}}
\safemath{\setW}{\mathcal{W}}
\safemath{\setX}{\mathcal{X}}
\safemath{\setY}{\mathcal{Y}}
\safemath{\setZ}{\mathcal{Z}}
\safemath{\emptySet}{\varnothing}

\safemath{\colA}{\mathscr{A}}
\safemath{\colB}{\mathscr{B}}
\safemath{\colC}{\mathscr{C}}
\safemath{\colD}{\mathscr{D}}
\safemath{\colE}{\mathscr{E}}
\safemath{\colF}{\mathscr{F}}
\safemath{\colG}{\mathscr{G}}
\safemath{\colH}{\mathscr{H}}
\safemath{\colI}{\mathscr{I}}
\safemath{\colJ}{\mathscr{J}}
\safemath{\colK}{\mathscr{K}}
\safemath{\colL}{\mathscr{L}}
\safemath{\colM}{\mathscr{M}}
\safemath{\colN}{\mathscr{N}}
\safemath{\colO}{\mathscr{O}}
\safemath{\colP}{\mathscr{P}}
\safemath{\colQ}{\mathscr{Q}}
\safemath{\colR}{\mathscr{R}}
\safemath{\colS}{\mathscr{S}}
\safemath{\colT}{\mathscr{T}}
\safemath{\colU}{\mathscr{U}}
\safemath{\colV}{\mathscr{V}}
\safemath{\colW}{\mathscr{W}}
\safemath{\colX}{\mathscr{X}}
\safemath{\colY}{\mathscr{Y}}
\safemath{\colZ}{\mathscr{Z}}

\safemath{\opA}{\mathbb{A}}
\safemath{\opB}{\mathbb{B}}
\safemath{\opC}{\mathbb{C}}
\safemath{\opD}{\mathbb{D}}
\safemath{\opE}{\mathbb{E}}
\safemath{\opF}{\mathbb{F}}
\safemath{\opG}{\mathbb{G}}
\safemath{\opH}{\mathbb{H}}
\safemath{\opI}{\mathbb{I}}
\safemath{\opJ}{\mathbb{J}}
\safemath{\opK}{\mathbb{K}}
\safemath{\opL}{\mathbb{L}}
\safemath{\opM}{\mathbb{M}}
\safemath{\opN}{\mathbb{N}}
\safemath{\opO}{\mathbb{O}}
\safemath{\opP}{\mathbb{P}}
\safemath{\opQ}{\mathbb{Q}}
\safemath{\opR}{\mathbb{R}}
\safemath{\opS}{\mathbb{S}}
\safemath{\opT}{\mathbb{T}}
\safemath{\opU}{\mathbb{U}}
\safemath{\opV}{\mathbb{V}}
\safemath{\opW}{\mathbb{W}}
\safemath{\opX}{\mathbb{X}}
\safemath{\opY}{\mathbb{Y}}
\safemath{\opZ}{\mathbb{Z}}
\safemath{\opZero}{\mathbb{O}}
\safemath{\identityop}{\opI}

\safemath{\veca}{\bma}
\safemath{\vecb}{\bmb}
\safemath{\vecc}{\bmc}
\safemath{\vecd}{\bmd}
\safemath{\vece}{\bme}
\safemath{\vecf}{\bmf}
\safemath{\vecg}{\bmg}
\safemath{\vech}{\bmh}
\safemath{\veci}{\bmi}
\safemath{\vecj}{\bmj}
\safemath{\veck}{\bmk}
\safemath{\vecl}{\bml}
\safemath{\vecm}{\bmm}
\safemath{\vecn}{\bmn}
\safemath{\veco}{\bmo}
\safemath{\vecp}{\bmp}
\safemath{\vecq}{\bmq}
\safemath{\vecr}{\bmr}
\safemath{\vecs}{\bms}
\safemath{\vect}{\bmt}
\safemath{\vecu}{\bmu}
\safemath{\vecv}{\bmv}
\safemath{\vecw}{\bmw}
\safemath{\vecx}{\bmx}
\safemath{\vecy}{\bmy}
\safemath{\vecz}{\bmz}

\safemath{\veczero}{\bmzero}
\safemath{\vecone}{\bmone}
\safemath{\vecxi}{\bmxi}
\safemath{\veclambda}{\bmlambda}
\safemath{\vecmu}{\bmmu}
\safemath{\vectheta}{\bmtheta}
\safemath{\vecphi}{\bmphi}
\safemath{\vecdelta}{\bmdelta}

\safemath{\matA}{\bA}
\safemath{\matB}{\bB}
\safemath{\matC}{\bC}
\safemath{\matD}{\bD}
\safemath{\matE}{\bE}
\safemath{\matF}{\bF}
\safemath{\matG}{\bG}
\safemath{\matH}{\bH}
\safemath{\matI}{\bI}
\safemath{\matJ}{\bJ}
\safemath{\matK}{\bK}
\safemath{\matL}{\bL}
\safemath{\matM}{\bM}
\safemath{\matN}{\bN}
\safemath{\matO}{\bO}
\safemath{\matP}{\bP}
\safemath{\matQ}{\bQ}
\safemath{\matR}{\bR}
\safemath{\matS}{\bS}
\safemath{\matT}{\bT}
\safemath{\matU}{\bU}
\safemath{\matV}{\bV}
\safemath{\matW}{\bW}
\safemath{\matX}{\bX}
\safemath{\matY}{\bY}
\safemath{\matZ}{\bZ}
\safemath{\matzero}{\bmzero}

\safemath{\matDelta}{\bDelta}
\safemath{\matLambda}{\bLambda}
\safemath{\matPhi}{\bPhi}
\safemath{\matSigma}{\bSigma}
\safemath{\matOmega}{\bOmega}
\safemath{\matTheta}{\bTheta}

\safemath{\matidentity}{\matI}
\safemath{\matone}{\matO}

\safemath{\rnda}{A}
\safemath{\rndb}{B}
\safemath{\rndc}{C}
\safemath{\rndd}{D}
\safemath{\rnde}{E}
\safemath{\rndf}{F}
\safemath{\rndg}{G}
\safemath{\rndh}{H}
\safemath{\rndi}{I}
\safemath{\rndj}{J}
\safemath{\rndk}{K}
\safemath{\rndl}{L}
\safemath{\rndm}{M}
\safemath{\rndn}{N}
\safemath{\rndo}{O}
\safemath{\rndp}{P}
\safemath{\rndq}{Q}
\safemath{\rndr}{R}
\safemath{\rnds}{S}
\safemath{\rndt}{T}
\safemath{\rndu}{U}
\safemath{\rndv}{V}
\safemath{\rndw}{W}
\safemath{\rndx}{X}
\safemath{\rndy}{Y}
\safemath{\rndz}{Z}

\safemath{\rveca}{\bimA}
\safemath{\rvecb}{\bimB}
\safemath{\rvecc}{\bimC}
\safemath{\rvecd}{\bimD}
\safemath{\rvece}{\bimE}
\safemath{\rvecf}{\bimF}
\safemath{\rvecg}{\bimG}
\safemath{\rvech}{\bimH}
\safemath{\rveci}{\bimI}
\safemath{\rvecj}{\bimJ}
\safemath{\rveck}{\bimK}
\safemath{\rvecl}{\bimL}
\safemath{\rvecm}{\bimM}
\safemath{\rvecn}{\bimN}
\safemath{\rveco}{\bomO}
\safemath{\rvecp}{\bimP}
\safemath{\rvecq}{\bimQ}
\safemath{\rvecr}{\bimR}
\safemath{\rvecs}{\bimS}
\safemath{\rvect}{\bimT}
\safemath{\rvecu}{\bimU}
\safemath{\rvecv}{\bimV}
\safemath{\rvecw}{\bimW}
\safemath{\rvecx}{\bimX}
\safemath{\rvecy}{\bimY}
\safemath{\rvecz}{\bimZ}

\safemath{\rvecxi}{\bmxi}
\safemath{\rveclambda}{\bmlambda}
\safemath{\rvecmu}{\bmmu}
\safemath{\rvectheta}{\bmtheta}
\safemath{\rvecphi}{\bmphi}

\safemath{\rmatA}{\bimA}
\safemath{\rmatB}{\bimB}
\safemath{\rmatC}{\bimC}
\safemath{\rmatD}{\bimD}
\safemath{\rmatE}{\bimE}
\safemath{\rmatF}{\bimF}
\safemath{\rmatG}{\bimG}
\safemath{\rmatH}{\bimH}
\safemath{\rmatI}{\bimI}
\safemath{\rmatJ}{\bimJ}
\safemath{\rmatK}{\bimK}
\safemath{\rmatL}{\bimL}
\safemath{\rmatM}{\bimM}
\safemath{\rmatN}{\bimN}
\safemath{\rmatO}{\bimO}
\safemath{\rmatP}{\bimP}
\safemath{\rmatQ}{\bimQ}
\safemath{\rmatR}{\bimR}
\safemath{\rmatS}{\bimS}
\safemath{\rmatT}{\bimT}
\safemath{\rmatU}{\bimU}
\safemath{\rmatV}{\bimV}
\safemath{\rmatW}{\bimW}
\safemath{\rmatX}{\bimX}
\safemath{\rmatY}{\bimY}
\safemath{\rmatZ}{\bimZ}

\safemath{\rmatDelta}{\bimDelta}
\safemath{\rmatLambda}{\bimLambda}
\safemath{\rmatPhi}{\bimPhi}
\safemath{\rmatSigma}{\bimSigma}
\safemath{\rmatOmega}{\bimOmega}
\safemath{\rmatTheta}{\bimTheta}

%% file: macros/standard-macros.tex
\usepackage{amssymb}
\usepackage{amsfonts}
\usepackage{mathrsfs}
\usepackage{xspace}
\usepackage{bm}
\usepackage{fancyref}
\usepackage{textcomp}

\usepackage{multirow}
\usepackage{stmaryrd}

\newenvironment{textbmatrix}{	\setlength{\arraycolsep}{2.5pt}%
								\big[\begin{matrix}}{\end{matrix}\big]%
								\raisebox{0.08ex}{\vphantom{M}}}

\def\be{\begin{equation}}
\def\ee{\end{equation}}
\def\een{\nonumber \end{equation}}
\def\mat{\begin{bmatrix}}
\def\emat{\end{bmatrix}}
\def\btm{\begin{textbmatrix}}
\def\etm{\end{textbmatrix}}

\def\ba#1\ea{\begin{align}#1\end{align}}
\def\bas#1\eas{\begin{align*}#1\end{align*}}
\def\bs#1\es{\begin{split}#1\end{split}} 
\def\bg#1\eg{\begin{gather}#1\end{gather}}
\def\bml#1\eml{\begin{multline}#1\end{multline}}
\def\bi#1\ei{\begin{itemize}#1\end{itemize}}

\safemath{\dirac}{\delta}					
\safemath{\krond}{\dirac}					

\safemath{\upto}{\uparrow}
\safemath{\downto}{\downarrow}
\safemath{\iu}{j}							
\safemath{\ev}{\lambda}						
\safemath{\hilseqspace}{l^{2}}				
\newcommand{\banachfunspace}[1]{\setL^{#1}}	
\safemath{\hilfunspace}{\banachfunspace{2}}	

\safemath{\SNR}{\textsf{SNR}} 				
\safemath{\PAR}{\textsf{PAR}} 				
\safemath{\No}{N_0}							
\safemath{\Es}{E_s}							
\safemath{\Eb}{E_b}							
\safemath{\EbNo}{\frac{\Eb}{\No}}
\safemath{\EsNo}{\frac{\Es}{\No}}

\DeclareMathOperator{\CHop}{\ensuremath{\opH}} 
\safemath{\tvir}{\rndh_{\CHop}}				
\safemath{\tvtf}{\rndl_{\CHop}}				
\safemath{\spf}{\rnds_{\CHop}}				
\safemath{\bff}{H_{\CHop}}					

\safemath{\ircf}{r_{h}}						
\safemath{\tftvcf}{r_{s}}					
\safemath{\tfcf}{r_{l}}						
\safemath{\bfcf}{r_{H}}						

\safemath{\tcorr}{c_h}						
\safemath{\scf}{c_{s}}						
\safemath{\tfcorr}{c_{l}}					
\safemath{\fcorr}{c_{H}}						

\safemath{\mi}{I}							
\safemath{\capacity}{C}						

\safemath{\normal}{\mathcal{N}}			
\safemath{\jpg}{\mathcal{CN}}			
\safemath{\mchain}{\leftrightarrow}		

\safemath{\dB}{\,\mathrm{dB}}
\safemath{\dBm}{\,\mathrm{dBm}}
\safemath{\Hz}{\,\mathrm{Hz}}
\safemath{\kHz}{\,\mathrm{kHz}}
\safemath{\MHz}{\,\mathrm{MHz}}
\safemath{\GHz}{\,\mathrm{GHz}}
\safemath{\s}{\,\mathrm{s}}
\safemath{\ms}{\,\mathrm{ms}}
\safemath{\mus}{\,\mathrm{\text{\textmu}s}}
\safemath{\ns}{\,\mathrm{ns}}
\safemath{\ps}{\,\mathrm{ps}}
\safemath{\meter}{\,\mathrm{m}}
\safemath{\mm}{\,\mathrm{mm}}
\safemath{\cm}{\,\mathrm{cm}}
\safemath{\m}{\,\mathrm{m}}
\safemath{\W}{\,\mathrm{W}}
\safemath{\mW}{\, \mathrm{mW}}
\safemath{\J}{\,\mathrm{J}}
\safemath{\K}{\,\mathrm{K}}
\safemath{\bit}{\,\mathrm{bit}}
\safemath{\nat}{\,\mathrm{nat}}

\safemath{\define}{\triangleq}			

\safemath{\equivalent}{\sim}
\safemath{\distas}{\sim}					
\safemath{\sdiff}{\Delta}				

\safemath{\reals}{\mathbb{R}}
\safemath{\positivereals}{\reals_{+}}
\safemath{\integers}{\mathbb{Z}}
\safemath{\posint}{\integers_{+}}
\safemath{\naturals}{\mathbb{N}}
\safemath{\posnaturals}{\naturals_{+}}
\safemath{\complexset}{\mathbb{C}}
\safemath{\rationals}{\mathbb{Q}}

\newcommand*{\fancyrefapplabelprefix}{app}		
\newcommand*{\fancyrefthmlabelprefix}{thm}		
\newcommand*{\fancyreflemlabelprefix}{lem}		
\newcommand*{\fancyrefcorlabelprefix}{cor}		
\newcommand*{\fancyrefdeflabelprefix}{def}		
\newcommand*{\fancyrefproplabelprefix}{prop}	
\newcommand*{\fancyrefobslabelprefix}{obs}		
\newcommand*{\fancyrefalglabelprefix}{alg}		
\newcommand*{\fancyrefasmlabelprefix}{asm}	    
\newcommand*{\fancyreftbllabelprefix}{tbl}	    

\frefformat{vario}{\fancyrefseclabelprefix}{Sec.~#1}
\frefformat{vario}{\fancyrefthmlabelprefix}{Thm.~#1}
\frefformat{vario}{\fancyreflemlabelprefix}{Lem.~#1}
\frefformat{vario}{\fancyrefcorlabelprefix}{Corr.~#1}
\frefformat{vario}{\fancyrefdeflabelprefix}{Def.~#1}
\frefformat{vario}{\fancyrefobslabelprefix}{Obs.~#1}
\frefformat{vario}{\fancyrefasmlabelprefix}{Ass.~#1}
\frefformat{vario}{\fancyreffiglabelprefix}{Fig.~#1}
\frefformat{vario}{\fancyrefapplabelprefix}{App.~#1} 
\frefformat{vario}{\fancyrefproplabelprefix}{Prop.~#1}
\frefformat{vario}{\fancyrefalglabelprefix}{Alg.~#1}
\frefformat{vario}{\fancyrefeqlabelprefix}{(#1)}
\frefformat{vario}{\fancyreftbllabelprefix}{Table~#1}

%% file: macros/defs.tex
\safemath{\dictab}{[\,\dicta\,\,\dictb\,]}

\safemath{\ysig}{\bmy}
\safemath{\ysighat}{\hat{\ysig}}
\safemath{\ysigdim}{M}
\safemath{\xsig}{\bmx}
\safemath{\xsigdim}{N}
\safemath{\nx}{n_x}
\safemath{\zsig}{\bmz}
\safemath{\zsigdim}{\ysigdim}
\safemath{\rsig}{\bmr}
\safemath{\Adict}{\bA}
\safemath{\Adicttilde}{\widetilde{\Adict}}
\safemath{\Adictdim}{\outputdim\times\xsigdim}
\safemath{\avec}{\bma}
\safemath{\avectilde}{\tilde{\avec}}
\safemath{\Bdict}{\bB}
\safemath{\Bdicttilde}{\widetilde{\Bdict}}
\safemath{\Cdict}{\bC}
\safemath{\cvec}{\bmc}
\safemath{\Ddict}{\bD}
\safemath{\Ddictdim}{\ysigdim\times\xsigdim}
\safemath{\dvec}{\bmd}
\safemath{\Ddicttilde}{\widetilde{\bD}}
\safemath{\Bonb}{\bB}
\safemath{\bvec}{\bmb}
\safemath{\Bonbdim}{\ysigdim\times\ysigdim}
\safemath{\noise}{\bmn}
\safemath{\noisedim}{\ysigim}
\safemath{\err}{\bme}
\safemath{\errdim}{\ysigdim}
\safemath{\errset}{\setE}
\safemath{\nerr}{n_e}
\safemath{\delop}{\bP_\errset}
\safemath{\delopc}{\bP_{{\errset}^c}}

\safemath{\cplxi}{\imath}
\safemath{\cplxj}{\jmath}

\safemath{\dict}{\matD}
\safemath{\inputdim}{N}		
\safemath{\outputdim}{M}		
\safemath{\sparsity}{S}	
\safemath{\inputdimA}{{N_a}}	
\safemath{\inputdimB}{{N_b}}	
\safemath{\elemA}{{n_a}}	
\safemath{\elemB}{{n_b}}	
\safemath{\resA}{\matR_a}	
\safemath{\resB}{\matR_b}	
\safemath{\subD}{\matS} 
\safemath{\subA}{\matS_a} 
\safemath{\subB}{\matS_b} 
\safemath{\dicta}{\matA} 	
\safemath{\dictb}{\matB} 	
\safemath{\hollowS}{H}
\safemath{\hollowA}{H_a}
\safemath{\hollowB}{H_b}
\safemath{\cross}{Z}
\safemath{\coh}{\mu_d}			
\safemath{\coha}{\mu_a}			
\safemath{\cohb}{\mu_b}			
\safemath{\mubs}{\nu}	
\safemath{\cohm}{\mu_m} 
\safemath{\dictset}{\setD}	
\safemath{\dictsetp}{\dictset(\coh,\coha,\cohb)}	
\safemath{\dictsetgen}{\dictset_\text{gen}}
\safemath{\dictsetgenp}{\dictsetgen(\coh)}
\safemath{\dictsetonb}{\dictset_\text{onb}}
\safemath{\dictsetonbp}{\dictsetonb(\coh)}

\safemath{\leftside}{U}
\safemath{\rightsideA}{R_a}
\safemath{\rightsideB}{R_b}

\safemath{\indexS}{\setI_S} 

\safemath{\na}{n_a}			
\safemath{\nb}{n_b}			
\safemath{\coeffa}{p_i}	
\safemath{\coeffb}{q_j}	
\safemath{\seta}{\setP}		
\safemath{\setb}{\setQ}     
\safemath{\setw}{\setW}	
\safemath{\setz}{\setZ}	
\safemath{\cola}{\veca}		
\safemath{\colb}{\vecb}		
\safemath{\cold}{\vecd}		
\safemath{\inputvec}{\vecx} 	
\safemath{\error}{\vece}	
\safemath{\noiseout}{\vecz} 	
\safemath{\inputvecel}{x}
\safemath{\inputveca}{\vecx_a}
\safemath{\inputvecb}{\vecx_b}
\safemath{\outputvec}{\vecy}	
\safemath{\lambdamin}{\lambda_{\mathrm{min}}}

\safemath{\elltwo}{\ell_2}
\safemath{\ellone}{\ell_1}
\safemath{\ellzero}{\ell_0}
\safemath{\ellinf}{\ell_\infty}
\safemath{\ellinftilde}{\ell_{\widetilde\infty}}
\safemath{\licard}{Z(\coh,\coha,\cohb)}
\safemath{\xsol}{\hat{x}}
\safemath{\xbord}{x_b}		
\safemath{\xstat}{x_s}		
\safemath{\xstatLone}{\tilde{x}_s}
\safemath{\order}{\mathcal{O}} 
\safemath{\scales}{\Theta} 
\safemath{\ones}{\mathbf{1}} 
\safemath{\zeroes}{\mathbf{0}} 
\safemath{\thlone}{\kappa(\coh,\cohb)} 
\safemath{\constoneA}{\delta} 
\safemath{\constoneB}{\epsilon} 
\safemath{\nlarge}{L}				   
\safemath{\sumlarge}{S_\nlarge}
\safemath{\maxlarger}{P_\nlarge}	   
\safemath{\Pzero}{\textrm{P0}}	
\safemath{\Pone}{\textrm{P1}}
\safemath{\vecfir}{\vecw}			 
\safemath{\vecsec}{\vecz}
\safemath{\elvecfir}{w}              
\safemath{\elvecsec}{z}				 
\safemath{\nlargefir}{n}
\safemath{\normout}{\gamma}
\safemath{\auxfun}{h}
\safemath{\supp}{\textrm{supp}}

\safemath{\indexa}{\ell}
\safemath{\indexb}{r}
\safemath{\indexc}{i}
\safemath{\indexd}{j}

\safemath{\project}{P}